**Superconducting Nanowire Single-Photon Detectors and effect of accumulation and unsteady releases of excess energy in materials.**

S .Pereverzev, G. Carosi, V. Li
Lawrence Livermore National Laboratory, California, USA
LLNL-JRNL-833208-DRAFT

*Executive summary*

*SNSPD is currently the fastest, most sensitive, and most efficient single-photon detector technology for near-IR to mid-IR range (1-15 µm), with the lowest dark counts and the highest (among superconducting photon detectors) operation temperature.*

1. *We suggest improving the SNSPD energy threshold to enable new applications in*
- *Astro-particles physics: low-energy neutrinos, CMB, dark matter particles, axions*
- *Space science: FIR-microwave imaging, spectroscopy, long-distance communications*
- *Metrology and spectral/chemical analysis, including chemical imaging of live cells*
- *Quantum information (QI): communication, quantum-conventional electronics interface*

2. *The low dark counts and absence of dissipation in SNSPD allow testing the hypothesis of excess noise generation by accumulation and unsteady energy releases in devices with energy flow, like the Self-Organized Criticality (SOC) dynamic, which can apply to a range of problems:*
- *Excess low-energy background in dark matter detectors*
- *Non-thermal noise and decoherence in quantum sensors*
- *Physics of glasses (disordered solids) and glass-like relaxation in QI devices.*

*An attempt to apply these ideas to SNSPD is risky, but the scientific and practical benefits of decreasing the energy threshold for photon detection would be enormously high. The common-sense arguments justify the suggested steps for energy threshold decrease and do not need SOC-like model applicability for superconducting devices. On the other hand, demonstrating the relevance of the SOC-like model to superconducting (quantum) sensors would have detrimental effects on neutrino and dark matter particle detectors and understanding noise and decoherence sources in quantum sensors.*

Interest in CEvNS, dark matter particles, and physics beyond the standard model results in many experiments in the limits of detectors' sensitivity and increasing body of data on intrinsic and extrinsic noise and low-energy background in different devices, including quantum sensors [1,2]. Systematization of noise and background data can accelerate detector progress and affect the development of new dark matter particle models, as potentially experimentally testable models are getting more attention.

We discuss noise generating mechanisms in the systems with energy flow, where thermal equilibrium approximation is not working, but solving exactly dynamical equations is currently impossible.

Ionizing radiation, mechanical deformation, thermo-mechanical stress can lead to energy accumulation in materials in the form of trapped charges, long-living excitations, defects, etc. Unsteady releases of the stored energy can lead to noise and excess background in the detector. Delayed relaxation events can follow any "shock": energetic particle, mechanical stress, etc. Interactions of excitations in materials (energy-bearing states and defects) can cause correlations in the stored energy releases and relaxational avalanches.

The Self-Organized Criticality theory is based on computer modeling of the dynamics of large ensembles of particles with known interactions [3,4]. Dynamical features observed in models are often observable in complex systems with energy flow [3,4], though the strict mathematical theory is lacking. For example, adding sand to the top of the sandpile results in sand avalanches on the sides, and the distribution of avalanches size follows SOC predictions.

Changes in temperature, pressure, mechanical stress, electric and magnetic field, number of defects, and impurities affect interactions between excitations (energy bearing states) in materials. Thus, environmental factors will affect relaxation events' properties (intensity, shape, spectrum). The time scale for different excitations/defects to store excess energy is wary for many orders of magnitude, and excitations and their interactions are material specific. Still, some features of the SOC model can be present:

- a spectrum of relaxation events (energy releases) polynomially decreasing with energy,
- noise power spectrum close to 1/f,
- increase in the number of relaxation events with energy influx,
- increase significant events rate when energy pumped into the system by small quanta
- suppression of significant relaxation events by facilitating quenching of excitations.

- *Important note: relaxation events can resemble interactions with particles but are not identical; an increase of energy and time resolution, using more information channels should help, though may require new technologies.*

Relaxation events can produce photons, phonons, ions, electrons, holes, and other types of quasiparticles. There are examples of energy transformation from one type to another- like dislocation or defect motion in metals, dielectrics, and semiconductors can result in luminescence, electron emission from the surface, spikes of conductivity (see ref. 4 for example). Some processes, like luminescence in electron-hole recombination, can be forbidden en masse in a given material. Still, the presence of defects and impurities can make low-intensity luminescence possible (silicon and porous silicon, for example). Thus, in looking for rare particle interactions, excluding rare chemical and condensed matter events could require careful experimental checks [2,6].

Recent progress in energy sensitivity and time resolution in solid-state low-temperature detectors (dark matter, CEvNS) allows new details of the energy relaxation process (seen before only in extraordinary experiments) [6]. Relaxation in semiconductors and dielectric singe crystal detectors after initial cool-down [1] resemble features observable in glasses (disordered solids): long, non-exponential, hysteretic, history-dependent; that can be re-started by slight temperature increase by several K (not to 300K) and cooling (see materials of the Excess workshop). Inelastic deformation (flow) of these crystals at a microscopic level goes through phase transitions in tiny fragments into different solid phases, sliding, formation of twins, dislocation formation, though micro-cracking (material braking) is also possible [7-10]. Thus, one can expect glass-like relaxation processes in micro-flow regions to appear due to thermomechanical stress during the cooling of joined dislike materials.

We can reverse argumentation here by assuming that glass-like relaxation processes indicate energy accumulation (and, likely, unsteady releases) occurs in material. With cooling to low and ultra-low (below 1mK) temperatures, glass-like properties (like memory effects) became more pronounced [11]. Subsystems in materials, like magnetic moments and spins in a superconductor, undergo a transition into a glass-like state with cooling [12]. We can also cite magnetic flux noise of SQIDs [13]; relaxation and charge noise in single-electron transition after applying a voltage step to the backside of the substrate [14] also resemble glass behavior. Thus, we argue here that changing electric or magnetic fields at low temperatures or applying AC EM fields or leaking of microwave or FR signals from the hot environment leads to energy accumulation in materials at low temperatures. Uneven releases of excess energy (pumped in by different mechanisms) stored in materials can lead to non-thermal noise and decoherence in quantum sensors and qubits.

We can re-phrase our hypothesis: the more energy is pumped into the material by readout or control system, the more non-thermal noise in low-temperature sensor or qubits one will see.

Our hypothesis suggests an apparent correlation between power dissipated by readout systems in different superconducting photon counters and noise equivalent power/detectable photon energy - see tab.1.

CMB and IR photon detectors:
<<< AC field 'drive' intensity, dissipation  <<< Noise equivalent power   "energy sensitivity">>>

| MKIDs | TES array with SQUID array readout | Superconducting Nanowire Single-Photon Detector |
|---|---|---|
| Sensors are integrated into microwave resonant circuits | TESs are DC-coupled to an array of SQUIDs. Frequency multiplexing readout | DC supercurrent in sensors while waiting for "click." |
| Continuous dissipation in sensors by microwave readout signals | Some leakage of RF signals to sensors, dissipation by DC in TES on transition. | No energy dissipation by readout in sensor while waiting for signal/photon |

SNSPDs demonstrate the best characteristics and the fastest progress in the field, which speaks toward the significance of SOC-like dynamics in photon sensors. We expect further improvements of SNSPD sensitivity and see new possibilities to study SOC-like dynamics in quantum devices using SNSPDs, as energy dissipation by the readout and, consequently, SOC-like dynamics are suppressed in these devices.

**Proposed avenue of study**

Superconducting Nanowire Single-Photon Detectors (SNSPDs) have demonstrated remarkable progress in recent years. They are among the fastest photon counters (~10 picosecond time resolution and ~1ns recovery time), have a low energy detection threshold (10 μm wavelength photon detection), and have low dark count rates. The simplified picture for an SNSPD operation is breaking a Cooper-pair by a photon and subsequent braking more Cooper-pairs by producing hot quasiparticles. This initial heating breaks superconductivity in the microscopic nanowire section and propagates the normal region along the nanowire due to heat produced by a small applied current. After the voltage pulse on the nanowire is detected, the small DC is switched off to let the nanowire cool down below the superconducting transition. Then the small current is applied again, and the device is ready to detect the next photon.

While there has been tremendous work building SNSPD arrays with high (but not the lowest possible) energy sensitivity of individual detectors, low-temperature experiments with SNSPD are still waiting. Many superconducting materials used for SNSPDs have low carrier concentrations, making it easier to change the electron concentrations in thin nanowires by an electric field. We will apply an electric field (perpendicular to the substrate) to lower the superconducting transition temperature in the wires, which will decrease the detector's energy threshold. Lowering the working temperature at the same time should help to keep low the number of temperature-activated counts.

When an SNSPD detector is waiting for an event (photon producing energetic quasiparticles in nanowire with superconducting current), no dissipation or energy production occurs in the detector materials, so SOC-type dynamic should be absent or strongly suppressed. We will apply low-intensity rf/microwave signals at frequencies below the superconducting gap, which will cause the production of sub-gap excitation in the materials but should not break the Cooper-pairs directly. Accumulation and correlated or avalanche-like relaxation of sub-gap excitations (i.e., SOC-like dynamics) can manifest as an increase in dark counts. By varying the frequency and intensity of applied RF signals, we could study both appearances of SOC dynamics, the density of sub-gap states, and the effects of populating these states on energy sensitivity and dark counts.

The application of small magnetic fields (both perpendicular and parallel to the nano-wires plane) is another avenue to study the microphysics of these detectors. Applying a parallel magnetic field can increase the transition temperature and critical current in ultra-thin films of some materials. Variation of electron concentration in the film/wire by electric field gives us an additional tool to study this puzzling effect.

All experiments mentioned above require good filtering and shielding to avoid uncontrollable exposure of samples to rf/microwave signals. Several enclosed shielding layers are required with microwave filers on all rf and DC lines at the entry into every next shield. To avoid leaking of rf/microwave noise from the input of the signal amplifier to the detector, filters or circulators are required, along with restricted bandwidth of the connecting line. Using quantum cascade lasers operating at 1-4 K temperature level to demonstrate photon sensitivity for the wavelength above 10 um is highly desirable but requires filtering and attenuation stages on optical fibers or waveguides coming to the cold detector at 10 mK. A combination of magnetic fields and superconducting shielding around samples lead to the experiment design with a small solenoid inside the superconducting shield.

Other applications.

The high energy sensitivity of SNSPD suggests the possibility of hot phonon detection, but we do not know about research in this area. Hot phonons detection on a large (kg-scale) single-crystal surface will allow new CEvNS and light dark matter particles detectors. A technique like phonon imaging could lead to directionality in these detectors.

Placing an IR/microwave photon detector inside the shield around quantum computer /qubits experiments can help check for hot photons leading to noise and decoherence. Hot photons can be leaking from outside or produced in materials in SOC-like processes due to microwave signals/ modulation applied to qubits.

"Smoking gun" evidence
There are contradictions in our current understanding of glasses.
 The tunneling two-level system model can quantitatively describe heat capacitance, ultrasound attenuation, and other properties of many glasses [15], but realistic microscopic models of TLS are absent [16]. As Anthony Leggett points out, other theories could be possible without using the TLS model. Experimental results incompatible with the TLS model are of high interest, which Anthony Leggett calls "smoking gun" evidence [17]. Demonstrating significant relaxational events caused by pumping energy into the material by small quanta could, potentially, provide such evidence.

References.

**Appendix. Tunneling Two-level Systems**

A dominant theoretical approach to non-thermal noise and decoherence in superconducting devices is based on a tunneling two-level systems model developed back in the 1970s to describe the properties of glasses (see, for example,[15]). The two-level systems (TLS) model postulates objects in glasses (in disordered solids) exist that are multi-particle configurations or single-particle /electron states with two closely arranged minimal energy configurations or closely spaced energy minima such that tunneling in between these configurations/states is possible. While TLS models can describe many features of glasses, the TLS model may not be the only explanation suited for describing the observed behaviors [17]. Microscopic models for TLS with required parameters (energy differences between the two adjacent minima and tunneling probabilities) are still absent [16].

At low temperatures, many subsystems in materials demonstrate complex and history-dependent relaxation properties. Glass-like relaxation processes mean that energy will be accumulated in materials due to the application of mechanical stress or any variations of electric or magnetic

field—due to signals intentionally applied to the cold device or leaking to the experiment from the hot environment. And as different non-equilibrium configurations of charges, magnetic moments, nuclear moments, defects, etc. are interacting directly and indirectly (through lattice deformations—i.e., via phonons—or electron systems—i.e., via Ruderman–Kittel–Kasuya–Yosida (RKKY) interactions), the authors posit avalanche-like relaxation events could also be possible. Thus, the SOC-type mechanism could be an addition to the TLS model for describing noise and decoherence in superconducting devices. There is an important difference though: in particle detectors, relaxation avalanches are mimicking low-energy interaction with an external particle, while in quantum devices, avalanches can lead to true energy-up conversion processes—like the creation of an energetic photon in a cold resonator or causing qubit transition into a high (well above temperature) energy state.